\documentclass[5p]{elsarticle}
\usepackage{amsmath,amssymb,color}
\usepackage{graphicx, graphics, hyperref}
\usepackage{psfrag}
\usepackage{textcomp}

\newcommand{\prd}{{Phys.~Rev.~D} }
\newcommand{\pre}{{Phys.~Rev.~E} }

\begin{document}

\title{R\'enyi entropy and the thermodynamic stability of black holes}

\author[nihon,wigner,cmat]{Viktor G.~Czinner}
\ead{czinner.viktor@wigner.mta.hu}
\author[nihon]{Hideo Iguchi}
\ead{iguchi.h@phys.ge.cst.nihon-u.ac.jp}
\address[nihon]{Laboratory of Physics, College of Science and Technology, Nihon University,
274-8501 Narashinodai, Funabashi, Chiba, Japan}
\address[wigner]{HAS Wigner Research Centre for Physics, H-1525 Budapest, P.O.~Box 49, Hungary}
\address[cmat]{Centro de Matem\'atica, Universidade do Minho, Campus de Gualtar, 
4710-057 Braga, Portugal}

\date{\today}

\begin{abstract}
Thermodynamic stability of black holes, described by the R\'enyi formula as equilibrium
compatible entropy function, is  investigated. It is shown that within this approach, 
asymptotically flat, Schwarzschild 
black holes can be in stable equilibrium with thermal radiation at a fixed temperature. 
This implies that the canonical ensemble exists just like in anti-de Sitter space, and 
nonextensive effects can stabilize the black holes in a very similar way as it is done 
by the gravitational potential of an anti-de Sitter space. Furthermore, it is also shown 
that a Hawking--Page-like black hole phase transition occurs at a critical temperature
which depends on the $q$-parameter of the R\'enyi formula.
  
\end{abstract}


\maketitle

\section{Introduction}
The aim of this Letter is to investigate the thermodynamic stability
problem of a Schwarzschild black hole based on a recent approach \cite{BC}, 
where the equilibrium compatible entropy function of the black hole is 
considered to be the R\'enyi one \cite{R1}.

The nonextensive nature of the Bekenstein-Hawking entropy of black hole 
event horizons has been noticed \cite{Davies} very early on after the 
thermodynamic theory of black holes had been formulated \cite{Bek},
and the corresponding thermodynamic stability problem has also been investigated 
many times with various approaches. 
The standard stability analysis of extensive systems however (with the criteria
that the {\it Hessian} of the entropy function has no positive eigenvalues), 
is not applicable for black holes, as it strongly depends on the 
{\it additive} property of the entropy function, which condition clearly fails 
to hold in this case.

The standard thermodynamic functions of a Schwarz\-schild black hole
are given by
\begin{equation}\label{st}
S_{BH}=4\pi M^2, \qquad 
\frac{1}{T_H}=\frac{\partial S_{BH}(M)}{\partial M}=8\pi M,
\end{equation}
and
\begin{equation}
C_{BH}=\frac{-S_{BH}'^2(M)}{S_{BH}''(M)}=-8\pi M^2 ,
\end{equation}
where $S_{BH}$ is the Bekenstein-Hawking entropy, $T_H$ is the Hawking
temperature and $C_{BH}$ is the corresponding heat capacity of the black hole.
In the classical approach (concluding from a {\it Hessian} 
analysis), Schwarzschild black holes appear to be thermodynamically unstable
in the canonical treatment, since the heat capacity of the hole is 
always negative. On the other hand, this
approach is clearly not reliable, as the Bekenstein-Hawking 
entropy is not additive, and the corresponding Hawking temperature 
is also not compatible with thermal equilibrium requirements 
\cite{BV}. For a better understanding on the problem, one needs to consider 
the consequences of nonadditive thermodynamic effects as well. 

To circumvent this issue, Kaburaki {\it et al}.~\cite{KOK1} have used
an alternative approach, and investigated the thermodynamic stability 
of black holes by the Poincar\'e turning point method \cite{Poincare}, 
which is a topological approach, and does not depend on the additivity 
of the entropy function. Later on, this method has been used to study 
critical phenomena of higher dimensional black holes and black rings as well
\cite{ALT}.

In \cite{Cz}, we investigated the Bekenstein-Hawking entropy problem of a
Schwarzschild black hole by considering the so-called formal logarithm approach 
\cite{BV} (discussed below), and found that (if the classical picture 
can be taken seriously without any quantum corrections in the small energy limit),
the equilibrium compatible entropy function of the black hole is 
linear in the hole's mass, and the corresponding zeroth law compatible 
temperature is constant, i.e.~it is independent of the hole's energy.
We also analyzed the thermodynamic stability of the problem, and showed
that isolated Schwarzschild black holes are stable against spherically 
symmetric perturbations within this approach. 

In the present Letter however, we are focusing on the direction 
that we proposed in \cite{BC}, where we regarded the Bekenstein-Hawking 
formula as a nonextensive Tsallis entropy \cite{T1}. This model was motivated
by the requirement of the existence of an empirical temperature in thermal 
equilibrium, or in other words, by the satisfaction of the zeroth law of 
thermodynamics. By applying the formal logarithm method \cite{BV}, we 
showed that the zeroth law compatible entropy function of black holes 
in this model is the R\'enyi one \cite{R1}, and the corresponding temperature 
function has an interesting similarity to the one of an AdS black hole in 
standard thermodynamics \cite{HP}. 

In the general case, both the Tsallis- and the R\'enyi entropies
contain a constant free parameter, whose physical meaning may depend 
on the concrete physical situation. In particular, for the problem of 
black hole thermodynamics, it may arise e.g.~from quantum corrections 
to micro black holes (a semi-classical approach has been obtained from 
the Bekenstein bound \cite{Bek2} in \cite{BC}), or from finite size 
reservoir corrections in the canonical ensemble \cite{B2,BBV}. Many 
other parametric situations are also possible.

The purpose of this Letter is to extend our study on the Tsallis-R\'enyi 
problem by investigating the 
corresponding thermodynamic stability of black holes. In the stability analysis 
we consider both the Poincar\'e turning point- and the {\it Hessian} methods 
because the R\'enyi entropy is additive for factorizing probabilities, and 
hence the standard  approach is also applicable. In the obtained results we 
find perfect agreement from both directions. Throughout the paper we use units 
such as $c=G=\hbar=k_B=1$.

\section{The Tsallis - R\'enyi approach} \label{nte}

Nonextensive approaches to black hole thermodynamics have been investigated
several times with various methods (see eg.~\cite{Lan1} and references therein), 
on the other hand, a zeroth law compatible formulation of nonextensive
thermodynamics is a long standing problem, and a possible solution 
has been proposed only very recently. 
Based only on the concept of composability, 
Abe showed \cite{Abe} that the most general nonadditive entropy composition rule 
which is compatible with homogeneous equilibrium has the form 
\begin{equation}\label{Abe}
H_{\lambda}(S_{12})=H_{\lambda}(S_1)+H_{\lambda}(S_2)+\lambda H_{\lambda}(S_1)H_{\lambda}(S_2), 
\end{equation}
where $H_{\lambda}$ is a differentiable function of $S$, $\lambda\in\mathbb{R}$ is a constant 
parameter,  and $S_{1}$, $S_{2}$ and $S_{12}$ are the entropies of the subsystems and the total 
system, respectively. 
By extending this result, Bir\'o and V\'an investigated non-homoge\-neous
systems as well \cite{BV}, and developed a formulation to determine the most general functional 
form of those nonadditive entropy composition rules that are compatible with the zeroth law of 
thermodynamics. They found that the general form is additive for the formal 
logarithms of the original quantities, which in turn, also satisfy the familiar 
relations of standard thermodynamics. They also showed, that for homogeneous systems  
the most general, zeroth law compatible entropy function has the form 
\begin{equation}\label{formlog}
L(S)=\frac{1}{\lambda}\ln[1+\lambda H_{\lambda}(S)],
\end{equation}
which is additive for composition, i.e.
\begin{equation}
L(S_{12})=L(S_{1})+L(S_{2}),
\end{equation}
and the corresponding zeroth law compatible temperature function can be obtained as
\begin{equation}
\frac{1}{T}=\frac{\partial L(S(E))}{\partial E}, 
\end{equation}
if one assumes additivity in the energy composition. 

For the classical black hole case, it is easy to show from the 
area law of the entropy function, that the Bekenstein-Hawking formula 
satisfies the equilibrium compatibility condition of (\ref{Abe}), as 
it follows the nonadditive composition rule
\begin{equation}\label{tc}
 S_{12}\ =\ S_1 +S_2 + 2\sqrt{S_1}\sqrt{S_2}\ , 
\end{equation}
which is equivalent with the choices of $H_{\lambda}(S)=\sqrt{S}$ and the $\lambda\rightarrow 0$ 
limit in the Abe formula \cite{Abe}. The corresponding thermodynamic and stability problem for the 
case of a Schwarzschild black hole (applying also the formal logarithm method) has been studied in 
\cite{Cz}. In the more general case however, when the parameter $\lambda \neq 0$, (originating e.g.~from
finite size reservoir corrections in the canonical approach \cite{B2,BBV}, or from  
quantum corrections to micro black holes (see e.g.~\cite{Carlip} and references therein)), 
the R\'enyi entropy formula may arise quite generally, when the conditions $L(0)=0$ and $L'(0)=1$ 
are also imposed, due to the consequence of some natural physical requirements (e.g.~triviality, 
and leading order additivity for small energies) \cite{BV}. 

The R\'enyi entropy \cite{R1}, defined as $S_R=\frac{1}{1-q}\ln\sum_ip^q_i$, 
is equivalent with the choices of 
$H_{\lambda}(S)=S$ and $\lambda=1-q$ in (\ref{formlog}), if the original entropy functions 
follow the nonadditive composition rule 
\begin{equation}\label{tr}
S_{12}=S_1+S_2+\lambda S_1S_2, 
\end{equation}
which is known as the Tsallis composition rule, and $q\in\mathbb{R}$ is the so-called 
nonextensivity parameter. The Tsallis entropy is defined as 
$S_T=\frac{1}{1-q}\sum_i(p^q_i-p_i)$ \cite{T1}, and it is easy to show that the formal 
logarithm of the Tsallis formula provides the R\'enyi entropy
\begin{equation}\label{ll}
S_R \equiv L(S_T)=\frac{1}{1-q}\ln\left[1+(1-q)S_T\right].
\end{equation}
In the limit of 
$q\rightarrow 1$ ($\lambda \rightarrow 0$), both the Tsallis- and the R\'enyi formulas reproduce the 
standard Boltzmann-Gibbs entropy, $S_{BG}=-\sum p_i\ln p_i$.

\section{Schwarzschild black holes}\label{sbh}

Based on the parametric Tsallis-R\'enyi model, we investigated 
the thermodynamic properties of a Schwarzschild black hole in \cite{BC}.
We found that the temperature--horizon radius relation is identical to 
the one obtained from a black hole in AdS space by using the original 
entropy formula, in both cases the temperature has a minimum. 
According to \eqref{ll} the R\'enyi entropy function of black holes 
can be obtained by taking the formal logarithm of the Bekenstein-Hawking 
entropy, which -- in the leading order of the $\lambda$ parameter --
follows the nonadditive Tsallis composition rule \eqref{tr}. Therefore,
the R\'enyi entropy of a black hole can be computed as 
\begin{equation}\label{sr}
S_R =\frac{1}{\lambda}\ln\left[1+\lambda S_{BH}\right],
\end{equation}
and for the Schwarzschild solution it results
\begin{eqnarray}\label{RENYI_SbhST}
&&S_R=\frac{1}{\lambda}\ln\left(1+4\pi \lambda M^2\right) ,\\
&&T_R=\frac{1}{8\pi M}+ \frac{\lambda}{2} M ,\quad 
C_R=\frac{8\pi M^2}{4\pi \lambda M^2-1}.
\end{eqnarray} 

\begin{figure}[!htb]
\noindent\hfil\includegraphics[scale=.3]{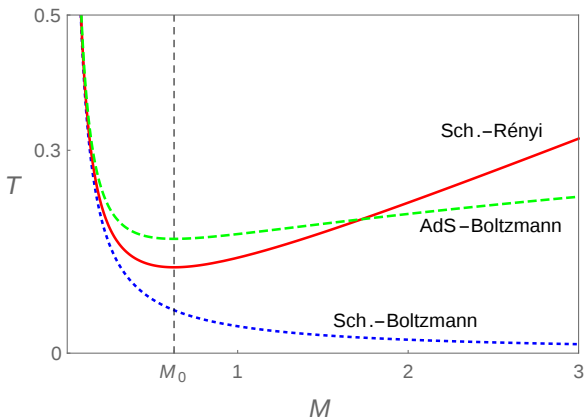} 
\caption{The figure shows the temperature of a Schwarzschild black hole
as a function of its mass-energy parameter in the asymptotically 
flat case with Boltzmann (blue, dotted) and R\'enyi (red, continuous) entropies, 
and also in AdS space with Boltzmann entropy (green, dashed).}
\label{fig:T}
\end{figure}

\begin{figure}[!htb]
\noindent\hfil\includegraphics[scale=.3]{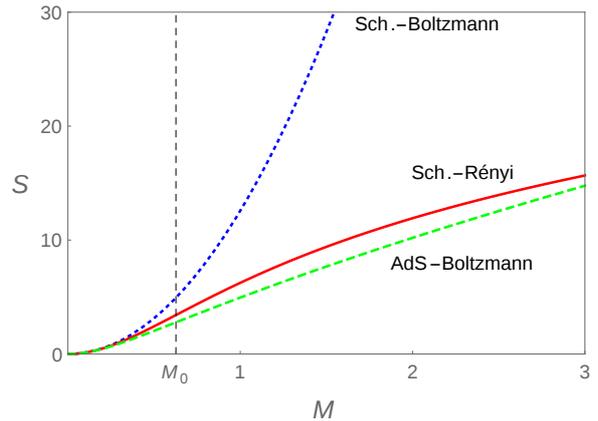} 
\caption{The figure shows the entropy of a Schwarzschild black hole
as a function of its mass-energy parameter in the asymptotically 
flat, Boltzmann (blue, dotted) and R\'enyi (red, continuous) cases, 
and also the Boltzmann case in AdS space (green, dashed).}
\label{fig:S}
\end{figure}

For comparison (not presented in \cite{BC}), on Fig.~\ref{fig:T} we plot the temperature functions 
versus the black hole mass for the Schwarzschild -- R\'enyi, the AdS -- Boltzmann and the 
standard Schwarzschild -- Boltzmann cases. On Fig.~\ref{fig:S} the corresponding entropy 
functions are  also plotted. On all plots in this paper we use the parameter value 
$\lambda=0.2$, and the corresponding curvature parameter of the AdS space is chosen 
such as to obtain the same $M_0$.

In the following section we discuss the thermodynamic stability of this
problem first by considering pure, isolated black holes in
the microcanonical treatment, and then by focusing on the canonical ensemble,
where the black holes are surrounded by a bath of thermal radiation.
For the purpose of generality, our analysis will be based on the Poincar\'e 
turning point method \cite{Poincare} following the works of Kaburaki et al.~\cite{KOK1}, 
however the canonical approach is also considered from the {\it Hessian} 
point of view. 

\section{Stability analysis}\label{s}

To separate stable and unstable configurations for cases of a one-parameter
series of equilibria, Poincar\'e developed a powerful analytic approach \cite{Poincare}. 
It has been applied several times to problems in astrophysical and gravitating
systems, in particular for the study of the thermodynamic stability
of black holes in standard four- \cite{KOK1} and also in higher dimensions
\cite{ALT}. Let us only quote here the main results of this method and omit all the details
and proofs which can be found in the original references.

Suppose $Z(x^i,y)$ is a distribution function whose extrema $\partial Z/\partial x^i=0$
define stable equilibrium configurations if the extremal value of $Z$ is a maximum. 
Consider now the equilibrium value $Z(y)=Z[X^i(y),y]$, where $X^i(y)$ is a solution of 
$\partial Z/\partial x^i=0$. If the derivative function $dZ/dy$, plotted versus $y$ has the 
topology of a continuous and differentiable curve, it can be shown that changes 
of stability will occur only at points where the tangents are vertical. The $Z$ 
distribution function is called {\it Massieu} function, the points 
with vertical tangents are called {\it turning points}, and the points with negative 
tangents near the turning points are a branch of {\it unstable} configurations, 
while the points with positive slopes near the turning points are a branch of 
{\it more stable} configurations.

Pure, isolated black holes are described by Kaburaki et al.~\cite{KOK1} when a perfectly 
reflecting, spherical mirror covers the hole just above its event horizon. In this 
idealistic case, the black hole can be described without radiation 
in the microcanonical treatment with $Z=Z(x^i,M)$ being the 
Bekenstein-Hawking entropy. A thermodynamic variable, 
$y$, is the total mass-energy, $M$, and the conjugate 
variable of $M$ with respect to the entropy is the derivative 
$\beta=\frac{\partial S}{\partial M}$, which is the inverse temperature. 

In the standard picture of black hole thermodynamics, $\beta$ is the inverse Hawking temperature, 
and the stability curve of a Schwarzschild black hole is the linear function $\beta(M)=8\pi M$. 
This straight line represents all the equilibrium 
configurations without any turning point, and hence the isolated 
Schwarzschild solution in vacuum is stable against spherically symmetric perturbations 
in the classical (microcanonical) treatment \cite{KOK1}. 

\begin{figure}[!htb]
\noindent\hfil\includegraphics[scale=.3]{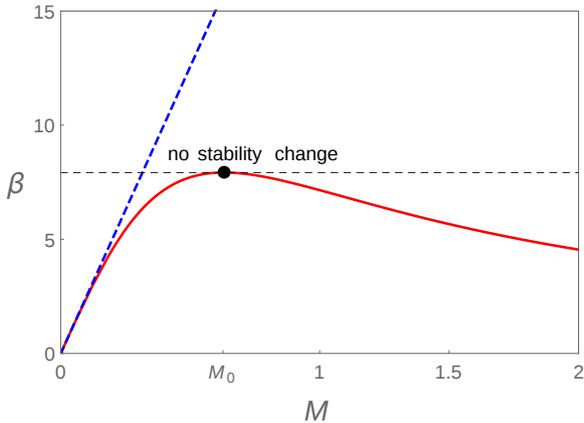} 
\caption{The figure shows the Poincar\'e stability curves of a Schwarzschild
black hole within the Boltzmann (blue, dashed) and the R\'enyi 
(red, continuous) approaches in the microcanonical treatment. No vertical 
tangent occurs in either case.}
\label{fig:Pmc}
\end{figure}

For the parametric R\'enyi case, the equilibrium compatible entropy function is given
in (\ref{RENYI_SbhST}), and $\beta$ is the inverse R\'enyi temperature
\begin{equation}
 \beta \equiv \frac{1}{T_R} = \frac{8\pi M}{1+4\pi \lambda M^2}\ .
\end{equation}
The corresponding stability curve $\beta(M)$ is plotted on Fig.~\ref{fig:Pmc} together
with the classical Schwarzschild curve for comparison.
As it can be seen, similarly to the standard result, 
the R\'enyi curve has no vertical tangent, and since this curve represents all 
the equilibrium configurations, we can conclude that isolated 
Schwarzschild black holes are thermodynamically stable against spherically symmetric 
perturbations (in the microcanonical treatment) in the R\'enyi approach as well.
 
When the black hole is surrounded by an infinite bath of thermal radiation, the 
appropriate thermodynamic approach to consider is the canonical one. In the traditional 
picture, the radiation is treated as an ideal reservoir with infinite size and an 
infinitely large heat capacity, so it can emit or absorb all the heat that is needed 
by any change of the black hole without modifying its temperature.
Recently, more realistic approaches have also been developed by considering large but
finite size reservoirs instead of the infinite approximation. In particular,
Bir\'o showed that finite size reservoir corrections can provide modifications 
to the standard canonical theory in the form of nonadditive thermodynamics \cite{B2}. 
These effects are usually neglected in the classical thermodynamic limit (when an infinite 
number of degrees of freedom is present), however they can provide relevant modifications 
to the behavior of finite size systems. 
In \cite{B2,BBV}, it has been shown, 
that in the case of a finite heat bath with a large but finite and constant heat capacity, 
the inverse heat capacity of the bath may serve as the $\lambda$ parameter 
in the equilibrium compatible entropy composition rule (\ref{Abe}), and provide the 
corresponding Tsallis- and R\'enyi entropy formulas in the leading order. 
These results provide additional motivation to investigate the R\'enyi approach 
for black hole thermodynamics in the canonical treatment. 

\begin{figure}[!htb]
\noindent\hfil\includegraphics[scale=.3]{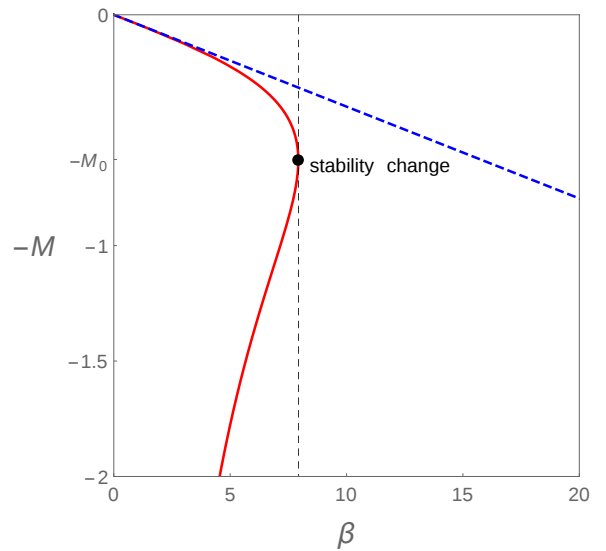} 
\caption{The figure shows the Poincar\'e stability curves of a Schwarzschild
black hole within the Boltzmann (blue, dashed) and the R\'enyi 
(red, continuous) approaches in the canonical treatment. For the R\'enyi curve, 
a vertical tangent occurs at $M_0$, which is the sign of a stability change.}
\label{fig:Pc}
\end{figure}

Let us now consider the black hole in the canonical approach. The black hole entropy 
is no longer the appropriate Massieu function, $Z$, which takes its maximum at a stable 
equilibrium, rather it is
\begin{equation}
 Z = S-\beta M \equiv -\beta F,
\end{equation}
where $F$ is the Helmholtz free energy \cite{KOK1}. The parameter $y$ now is $\beta$, 
and the conjugate variable, $dZ/dy$, is
\begin{equation}
 \frac{d(S-\beta M)}{d\beta} = -M .
\end{equation}
The corresponding stability curve is then $-M(\beta)$, which we plotted on Fig.~\ref{fig:Pc}.

The canonical stability curves are simply the $\pi/2$ clockwise rotated versions 
of the microcanonical ones, and it is immediate
to see that a vertical tangent appears at $M=M_0$, which belongs to the minimum temperature 
$T_0=\frac{1}{2}\sqrt{\frac{\lambda}{\pi}}$. Black holes with mass parameter
smaller than $M_0$ are {\it unstable} against spherically symmetric perturbations within 
this approach, however larger black holes with $M>M_0$ are {\it stable}, as opposed to 
the standard result where all solutions appear to be unstable.

\begin{figure}[!htb]
\noindent\hfil\includegraphics[scale=.3]{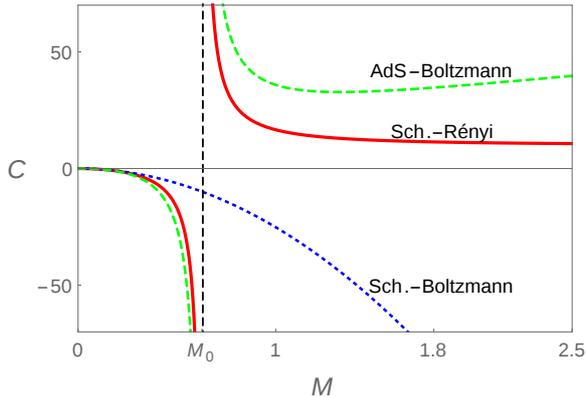} 
\caption{The figure shows the heat capacity of a Schwarzschild black hole
as a function of its mass-energy parameter in the standard (blue, dotted),
R\'enyi (red, continuous) and AdS (green, dashed) cases.}
\label{fig:C}
\end{figure}

The stability change at $M_0$ can also be confirmed by the {\it Hessian} analysis, i.e.~in 
this case simply by checking the signature of the heat capacity of the Schwarzschild black 
hole in the bath. The corresponding curves are plotted on Fig.~\ref{fig:C}.

It can be seen (as shown in \cite{BC}), that the heat capacity
has a pole at $M_0$, analogous to the AdS-Boltzmann case, and black holes with
larger mass parameter have positive heat capacities (i.e.~negative {\it Hessian}) 
and hence these solutions are thermodynamically stable. Black holes with smaller 
masses however are unstable solutions. 

\section{Phase Transition}\label{pt}
In their classic paper \cite{HP}, Hawking and Page investigated the thermodynamic
properties of black holes in asymptotically AdS spacetimes. They showed that due 
to the presence of the gravitational potential of the AdS space, the stability 
properties of black holes are different from their corresponding ones in asymptotically 
flat spacetimes. In particular, the canonical ensemble exists for Schwarzschild black holes, 
and they can be in stable equilibrium with thermal radiation at a fixed temperature.
In addition, it has been found that a thermodynamic phase transition occurs between 
the radiation and the black hole phases at a critical temperature which depends
on the AdS curvature parameter only.
Based on the similarity of our findings to the AdS problem, it is motivated to 
discuss the question of a possible phase transition in the Tsallis-R\'enyi 
approach as well.

By plotting the free energy of the black hole versus the temperature (for asymptotically 
flat black holes, i.e.~at zero pressure, the Gibbs free energy is equivalent with the Helmholtz 
one), a Hawking-Page-like transition can be revealed. 
\begin{figure}[!htb]
\noindent\hfil\includegraphics[scale=.33]{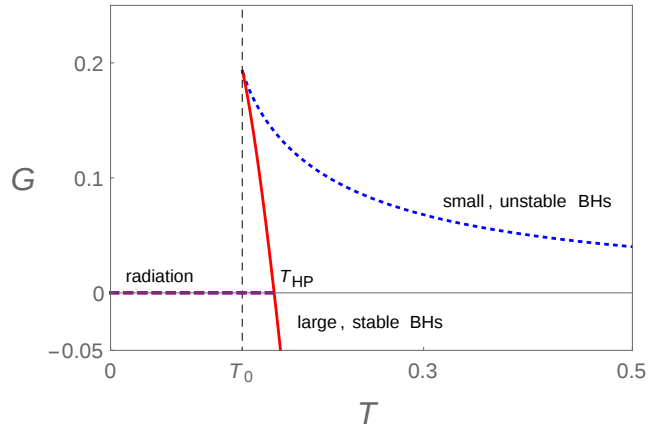} 
\caption{The figure shows the free energy of a Schwarzschild black hole
in the R\'enyi model as a function of the temperature in the canonical
treatment.}
\label{fig:HP}
\end{figure}
On Fig.~\ref{fig:HP}, the blue, dotted curve represents the branch of black holes with 
$M<M_0$, which are unstable at any temperature. Black holes with $M_0<M$ (red, continuous curve) 
are stable above the minimum temperature, but they posses positive free energy in the
$T_0\leq T<T_{HP}$ region, and therefore the pure radiation phase (purple, dashed line) 
is the thermodynamically preferred state with its approximately zero free energy. 
On the other hand, for $T_{HP}<T$, the large (and stable) black hole configuration
becomes the thermodynamically preferred state because its free energy is smaller
than zero, and hence, a phase transition occurs at $T_{HP}$ from the radiation- 
to the black hole phase.
From the $F\equiv M-TS=0$ condition we get $T_{HP}\approx 0.62 \sqrt{\frac{\lambda}{\pi}}$,
which depends on the $\lambda$ parameter only.

This picture is completely analogous to the Hawking-Page 
transition of Schwarzschild black holes in AdS space, where the curvature parameter plays 
a similar role that is played by the $\lambda$ ($\equiv 1-q$) parameter in the R\'enyi model.
A relevant difference however, compared to the AdS case, is that in our model the R\'enyi 
entropy is additive, and hence the corresponding stability results are reliable.

\section{Summary and discussion}

In this Letter we studied 
the thermodynamic stability problem of Schwarzschild black holes described by the R\'enyi
formula as equilibrium- and zeroth law compatible entropy function. First we considered
the question of a pure, isolated black hole in the microcanonical approach, and showed that
these configurations are stable against spherically symmetric perturbations, just like
in the classical picture. We also considered the problem when the black hole is surrounded 
by a bath of thermal radiation in the canonical treatment, and found that, 
as opposed to the standard picture, asymptotically flat, Schwarzschild black
holes can be in stable equilibrium with thermal radiation at a fixed temperature, and a 
stability change occurs at a certain value of the mass-energy parameter
which belongs to the minimum temperature solution. 
Black holes with smaller masses are unstable in this model, however larger black 
holes become stable. 
These results are very similar to the ones obtained by Hawking 
and Page in AdS space within the standard Boltzmann entropy approach \cite{HP}.
Based on this similarity, we also analysed the question of a
possible phase transition in the canonical picture, and showed that a 
Hawking-Page-like black hole phase transition occurs in a very similar 
fashion as in AdS space, and the corresponding critical temperature
depends only on the $q$-parameter of the R\'enyi formula.

Our findings are relevant in many aspects. Parametric corrections to 
the Bekenstein-Hawking entropy formula may arise in various kind of 
physical situations, most importantly from quantum considerations 
(stemming either from string theory, loop quantum gravity, or other 
semi-classical theories). In these corrections, the perturbation 
parameters are small, and by connecting them to the $\lambda$ parameter 
in Abe's formula (\ref{Abe}), it can be expected that the Tsallis 
composition rule (\ref{tc}) is obtained quite generally in the leading 
order. In addition, from the requirement of the existence of an empirical 
temperature in thermal equilibrium, the R\'enyi entropy arises very 
naturally via the formal logarithmic method \cite{BV}. Based on these 
lines, our obtained results seem to be quite generic for parametric 
corrections to the Bekenstein-Hawking model in the small parameter 
and small energy limit. 

As a different direction, we also mentioned that finite size reservoir 
corrections can result the same Tsallis-R\'enyi model in the canonical
picture \cite{B2,BBV}, and we expect that many other parametric situations 
are also possible. One of the motivations of this approach has been 
to satisfy the zeroth law of thermodynamics, and it is an important 
question whether the model is also in accordance with the remaining laws. 
In particular, Bekenstein's generalized second law \cite{BGSL} states 
that the sum of the black hole entropy and the common (ordinary) entropy 
in the black hole exterior never decreases on a statistical average. In 
obtaining this result, Bekenstein considered the Boltzmann-Gibbs formula 
for estimating the entropy of the system. Within our approach, Bekenstein's 
computations may be repeated by using the R\'enyi formula instead 
(see Sec.~\ref{nte} for the definition), and due to the properties of 
the R\'enyi entropy measure \cite{R1}, we expect that the generalized 
second law remains valid in this approach as well. We postpone this 
study for a future work. 

As for the third law, the question is even more open. Recent results 
suggest (see e.g.~\cite{Bagci} and references therein), that the 
generalized entropy formulas, in particular the R\'enyi entropy, 
violate the third law of thermodynamics even when the $q$-parameter 
is close to 1. The validity and applicability of the generalized 
entropies is therefore clearly an unsettled problem, and it is in 
the focus of active investigations today. There are many open questions 
in this field and also many research directions to consider. In this 
Letter, we studied the simplest problem of a standard 3+1 dimensional, 
Schwarzschild black hole, and our plan is to extend our work to more
general settings, e.g.~rotating, Kerr black holes, or dynamic black 
holes formed by collapsing shells. Due to its similarity to the AdS 
problem, our present results might have some relevance from the 
AdS/CFT correspondence \cite{MAL,WIT} point of view as well.
Further studies to address these questions are in progress and 
we hope to report on them in a forthcoming publication.

\section*{Acknowledgement}
V.G.Cz is grateful for discussions with Prof.~T.S.~Bir\'o. 
The research leading to this result has received funding from: the 
European Union Seventh Framework Programme (FP7/2007-2013) under 
the grant agreement No.~PCO\-FUND-GA-2009-246542; the FCT project 
SFRH/BCC/ 105835/2014; the Japanese Ministry of Education, Science,
Sports, and Culture Grant-in-Aid for Scientific Research (C) (No.~23540319),
and also from the Japan Society for the Promotion of Science L14710 grant.

\end{document}